\documentclass[twocolumn,nofootinbib,preprintnumbers,superscriptaddress,amsmath,amssymb,PRL]{revtex4-1}
\usepackage{graphicx}
\usepackage{gensymb}
\usepackage{dcolumn}
\usepackage{float}
\usepackage{mathptmx, courier, pifont}
\usepackage[scaled=0.92]{helvet}
\usepackage[T1]{fontenc}
\usepackage{textcomp}
\usepackage{color}
\usepackage[normalem]{ulem}

\usepackage[dvipsnames]{xcolor}
\usepackage[colorlinks=true,urlcolor=Blue,linkcolor=Blue]{hyperref}
\usepackage[all]{hypcap}
\usepackage{makecell}
\usepackage{multirow}
\usepackage{url}

\begin{document}



\title{Direct determination of the atomic mass difference of the pairs $^{76}$As-$^{76}$Se and $^{155}$Tb-$^{155}$Gd rules out $^{76}$As and $^{155}$Tb as possible candidates for electron (anti)neutrino mass measurements}
\author{Z.~Ge}\thanks{Corresponding author. Email address: z.ge@gsi.de}\thanks{Present address: GSI Helmholtzzentrum f\"ur Schwerionenforschung GmbH, 64291 Darmstadt, Germany}
\affiliation{Department of Physics, University of Jyv\"askyl\"a, P.O. Box 35, FI-40014, Jyv\"askyl\"a, Finland}%
\author{T.~Eronen}\thanks{Corresponding author. Email address: tommi.eronen@jyu.fi}
\affiliation{Department of Physics, University of Jyv\"askyl\"a, P.O. Box 35, FI-40014, Jyv\"askyl\"a, Finland}%
\author{A.~de Roubin}
\affiliation{Centre d'Etudes Nucl\'eaires de Bordeaux Gradignan, UMR 5797 CNRS/IN2P3 - Universit\'e de Bordeaux, 19 Chemin du Solarium, CS 10120, F-33175 Gradignan Cedex, France}%
\author{J.~Kostensalo}
\affiliation{Natural Resources Institute Finland, Yliopistokatu 6B, FI-80100, Joensuu, Finland}%
\author{J.~Suhonen}
\affiliation{Department of Physics, University of Jyv\"askyl\"a, P.O. Box 35, FI-40014, Jyv\"askyl\"a, Finland}%
\author{D.~A.~Nesterenko}
\affiliation{Department of Physics, University of Jyv\"askyl\"a, P.O. Box 35, FI-40014, Jyv\"askyl\"a, Finland}%
\author{O.~Beliuskina}
\affiliation{Department of Physics, University of Jyv\"askyl\"a, P.O. Box 35, FI-40014, Jyv\"askyl\"a, Finland}%
\author{R.~de~Groote}
\affiliation{Department of Physics, University of Jyv\"askyl\"a, P.O. Box 35, FI-40014, Jyv\"askyl\"a, Finland}%
\author{C.~Delafosse}
\affiliation{Department of Physics, University of Jyv\"askyl\"a, P.O. Box 35, FI-40014, Jyv\"askyl\"a, Finland}%
\author{S.~Geldhof}\thanks{Present address: KU Leuven, Instituut voor Kern- en Stralingsfysica, B-3001 Leuven, Belgium}
\affiliation{Department of Physics, University of Jyv\"askyl\"a, P.O. Box 35, FI-40014, Jyv\"askyl\"a, Finland}%
\author{W.~Gins}
\affiliation{Department of Physics, University of Jyv\"askyl\"a, P.O. Box 35, FI-40014, Jyv\"askyl\"a, Finland}%
\author{M.~Hukkanen}
\affiliation{Department of Physics, University of Jyv\"askyl\"a, P.O. Box 35, FI-40014, Jyv\"askyl\"a, Finland}%
\affiliation{Centre d'Etudes Nucl\'eaires de Bordeaux Gradignan, UMR 5797 CNRS/IN2P3 - Universit\'e de Bordeaux, 19 Chemin du Solarium, CS 10120, F-33175 Gradignan Cedex, France}
\author{A.~Jokinen} 
\affiliation{Department of Physics, University of Jyv\"askyl\"a, P.O. Box 35, FI-40014, Jyv\"askyl\"a, Finland}%
\author{A.~Kankainen}
\affiliation{Department of Physics, University of Jyv\"askyl\"a, P.O. Box 35, FI-40014, Jyv\"askyl\"a, Finland}%
\author{J.~Kotila}
\affiliation{Department of Physics, University of Jyv\"askyl\"a, P.O. Box 35, FI-40014, Jyv\"askyl\"a, Finland}%
\affiliation{Finnish Institute for Educational Research, University of Jyv\"askyl\"a, P.O. Box 35, FI-40014, Jyv\"askyl\"a, Finland}%
\affiliation{Center for Theoretical Physics, Sloane Physics Laboratory Yale University, New Haven, Connecticut 06520-8120, USA}
\author{\'A.~Koszor\'us}\thanks{Present address: Experimental Physics Department, CERN, CH-1211 Geneva 23, Switzerland}
\affiliation{Department of Physics, University of Liverpool, Liverpool, L69 7ZE,  United Kingdom}%
\author{I.~D.~Moore}
\affiliation{Department of Physics, University of Jyv\"askyl\"a, P.O. Box 35, FI-40014, Jyv\"askyl\"a, Finland}%
\author{A.~Raggio}
\affiliation{Department of Physics, University of Jyv\"askyl\"a, P.O. Box 35, FI-40014, Jyv\"askyl\"a, Finland}%
\author{S.~Rinta-Antila}
\affiliation{Department of Physics, University of Jyv\"askyl\"a, P.O. Box 35, FI-40014, Jyv\"askyl\"a, Finland}%
\author{V.~Virtanen}
\affiliation{Department of Physics, University of Jyv\"askyl\"a, P.O. Box 35, FI-40014, Jyv\"askyl\"a, Finland}%
\author{A.~P.~Weaver}
\affiliation{School of Computing, Engineering and Mathematics, University of Brighton, Brighton BN2 4JG, United Kingdom}%
\author{A.~Zadvornaya}\thanks{Present address: II. Physikalisches Institut, Justus-Liebig-Universit{\"a}t Gie{\ss}en, 35392 Gie{\ss}en, Germany}
\affiliation{Department of Physics, University of Jyv\"askyl\"a, P.O. Box 35, FI-40014, Jyv\"askyl\"a, Finland}
\date{\today}
\begin{abstract}
The first direct determination of the ground-state-to-ground-state $Q$ values of the $\beta^-$ decay $^{76}$As $\rightarrow$ $^{76}$Se and the electron-capture decay $^{155}$Tb $\rightarrow$ $^{155}$Gd was performed utilizing the double Penning trap mass spectrometer JYFLTRAP.  By measuring the atomic mass difference of the decay pairs via the  phase-imaging ion-cyclotron-resonance (PI-ICR) technique, the $Q$ values of $^{76}$As $\rightarrow$ $^{76}$Se and $^{155}$Tb $\rightarrow$ $^{155}$Gd were determined to be 2959.265(74) keV and 814.94(18) keV, respectively.
The precision was increased relative to earlier measurements by factors of 12 and 57, respectively. The new $Q$ values are  1.33 keV and 5 keV lower compared to the values adopted in the most recent Atomic Mass Evaluation 2020. 
With the newly determined ground-state-to-ground-state $Q$ values combined with the excitation energy from $\gamma$-ray spectroscopy, the $Q$ values for ground-state-to-excited-state transitions  $^{76}$As (ground state) $\rightarrow$ $^{76}$Se$^*$ (2968.4(7) keV) and  $^{155}$Tb (ground state) $\rightarrow$ $^{155}$Gd$^*$ (815.731(3) keV) were derived to be -9.13(70) keV and -0.79(18) keV.
Thus we have confirmed that both of the $\beta^{-}$-decay and EC-decay candidate transitions are energetically forbidden at a level of at least 4$\sigma$, thus definitely excluding these two cases from the list of potential candidates for the search of low-$Q$-value  $\beta^-$ or EC decays to determine the electron-(anti)neutrino mass.

\end{abstract}
\maketitle
\section{Introduction}

The knowledge of the absolute neutrino-mass scale is extremely important both for cosmological  models and for the fundamental understanding of the nature of particle masses. 
The information on  neutrino masses from oscillation experiments is  limited by the fact that these experiments are only able to  measure the differences of squared neutrino masses and not  their absolute mass scale. 
The only model-independent information on the neutrino masses, beyond just the mass differences, can be extracted from energy-momentum conservation relations in weak reactions like $\beta^-$ or electron-capture (EC) decay in which an antineutrino or a neutrino is involved~\cite{Suhonen1998,Avignone2008,Ejiri2019}. 
Enrico Fermi proposed in 1933 such a kinematic  search for the antineutrino mass $m_{\overline{\nu}_{e}}$ in the endpoint region of the $\beta^-$ spectra in $^{3}$H $\beta$ decay,  $^{3}$H(1/2$^{+}$) $\rightarrow$ $^{3}$He(1/2$^{+}$) + $e^-$ + $\overline{\nu}_e$, with a $Q$ value of of 18.59201(7) keV~\cite{Myers2015}. The 
KATRIN experiment
~\cite{aker2021direct}  sets an upper limit of  $m_{\overline{\nu}_{e}}$ < 0.8 eV, at 90\% Confidence Level (CL), via this decay thus improving the previous bound from the Mainz~\cite{Kraus2005} and Troitsk~\cite{Aseev2011} experiments which managed to constrain $m_{\overline{\nu}_{e}}$ < 2.2 eV at 95\% CL. KATRIN will continue running to reach a design sensitivity limit of $m_{\overline{\nu}_{e}}$ $\sim$ 0.2 eV after 5 calendar years of data taking. Project 8 is exploring a new technique for $\beta$ spectrometry based on cyclotron radiation utilizing molecular tritium as a source~\cite{Monreal2009}. 
In a similar manner as the $\beta^-$-decay end-point experiments, the sensitivity of EC experiments to the neutrino mass depends  on the fraction of events near the end-point. This fraction increases with decreasing $Q$ value  and the (anti)neutrino-mass sensitivity increases accordingly.


An alternative isotope to tritium is $^{163}$Ho which has the advantage of a small  $Q$ value of 2.833(30)$_{\rm stat}$(15)$_{\rm sys}$ keV~\cite{Eliseev2015}, decaying via EC to $^{163}$Dy. Currently, there are several next generation experiments, ECHo~\cite{Gastaldo2017}, HOLMES~\cite{Faverzani2016}, and NuMECS~\cite{Croce2016}, exploring this decay to probe the electron-neutrino mass.  These experiments are complementary to tritium-based searches from the perspective of technology.  In addition, the decay of $^{163}$Ho determines the electron-neutrino mass as opposed to the electron-antineutrino mass in tritium.

A low, especially ultra-low (< 1~keV), $Q$-value $\beta$ or EC decay is of great interest for possible future (anti)neutrino-mass determination experiments~\cite{Mustonen2010,Mustonen2011,Suhonen2014,DeRoubin2020,ge2021b}. If an excited state in the daughter nucleus can be found with an excitation energy close to the mass difference of the parent and decay-daughter atoms, which is equal to the ground-state-to-ground-state $Q$ value, the decay energy of the corresponding transition can be very low. 
The ultra-low $Q$-value decay  branch of $^{115}$In (9/2$^{+}$) to the first excited state of $^{115}$Sn$^*$ (9/2$^{+}$) was first revealed by Cattadori \emph{et al.}~\cite{Cattadori2005}.  
Two independent measurements of the $Q$ value of this branch, via Penning trap mass spectrometry (PTMS) at Florida State University~\cite{Mount2009} and at the University of Jyv\"askyl\"a~\cite{Wieslander2009}, confirmed it to be ultra-low. The $Q$ value of this branch was further refined to be 0.147(10) keV using the accurately measured excitation energy of the first excited state of $^{115}$Sn from~\cite{Zheltonozhsky_2018}.



The ground-state-to-ground-state decay $Q$ value ($Q_{0}$) for $\beta^-$ decay ($Q_{\beta^{-}}$) and EC decay ($Q_{EC}$) is the difference of atomic masses of the decay parent ($M_p$) and the decay daughter ($M_d$):
\begin{equation}
    Q_{0} = Q_{\beta^{-}/EC} = (M_p - M_d)c^2.
\end{equation}
where $c$ is the speed of light in vacuum.
Combining the  $Q_{0}$ value with the excitation energy $E{}{^*}$ of the decay-daughter state,  yields  the  ground-state-to-excited-state $Q$ value of $Q_{0}^{*}$:
\begin{equation}
Q_{0}^{*} = Q_{0}  -  E^*.
\end{equation}
For EC decay, the atomic binding energy $B_j$, where $j$ denotes the atomic shell of the captured electron, needs to be taken into account to derive the released energy in the decay:
\begin{equation}
Q_{EC,j}^{*} = Q_{EC}  -  E^* - B_j. 
\end{equation}

Two potential ultra-low $Q$ value candidate transitions, $\beta^-$ decay $^{76}$As (ground-state) $\rightarrow$ $^{76}$Se$^*$ (680.1035(17)~keV~\cite{NNDC}) and EC $^{155}$Tb (ground-state) $\rightarrow$ $^{155}$Gd$^*$ (2968.4(7)~keV~\cite{NNDC}), are of particular interest. Both of the transitions are possibly of the allowed type as shown in Table~\ref{table:low-Q1}.
Emphasis should be given to the transitions of allowed type for the direct (anti)neutrino mass determination~\cite{Suhonen2014,Gamage2019}.
They are promising candidates due to  two main reasons: first, they have larger branching ratios, enabling the accumulation of more data in a shorter time period and potentially making the case more lucrative  for direct (anti)neutrino mass determination; second, the decay transition is driven by decay matrix element(s) such that the beta spectral shape is universal, which makes the analysis of the $\beta$-decay spectrum nuclear-structure independent.
The evaluated $Q$ values in the Atomic Mass Evaluation 2020~\cite{Wang2021} (AME2020) and derived $Q_{0}^{*}$ value of the candidate transitions are tabulated in Table~\ref{table:low-Q1}.
The excitation energies $E^*$ in the daughter nuclei are well-known with sub-keV precision~\cite{NNDC} while the $Q_{0}$ values are poorly known and lack a value from a direct measurement. This results in the derived $Q_{{\beta^{-}/EC}}^{*}$ values of both decays a precision worse than 1 keV~\cite{Gamage2019,Wang2021}.  Currently, PTMS is the most precise and accurate method for determining atomic masses and $Q_{0}$ values, and it is capable of reaching a sub-keV precision in a direct manner. Hence, to confirm whether these two decays are energetically possible, 
direct PTMS determination of the $Q_{0}$ values is called for. In order to solve this puzzle, we have performed in this work a direct $Q_{0}$-value determination  for the two promising candidate nuclei, $^{76}$As and $^{155}$Tb, using the JYFLTRAP Penning-trap setup.

\begin{table*}[!htb]
   \caption[]{Potential ultra-low $Q$-value candidate transitions for electron-(anti)neutrino mass measurements: $\beta^-$ decay of the ground state of $^{76}$As to an excited state in $^{76}$Se and EC decay of the ground state of $^{155}$Tb to an excited state in $^{155}$Gd. The first column gives the initial ground state of the parent nucleus and the second column  gives the half-life of the parent nucleus.
 The third column gives  the excited final state of interest for the low-$Q$-value transition. The fourth column gives the decay type and the fifth column gives the $\beta^-$/EC-decay $Q_{0}$ value in units of keV, taken from literature (AME2020)~\cite{Wang2021}.  The sixth column gives the derived $Q^{*}_{0}$ value. The  last  column gives the experimental excitation energy E$^{*}$ (keV) with the experimental error~\cite{NNDC}. 
 {Spin-parity assignments enclosed by the braces indicate that these are uncertain, which results in an uncertainty in the decay type, indicated by a  \{?\}}.
 }
  \begin{ruledtabular}
   \begin{tabular*}{\textwidth}{lccccccc}
Initial state &$T_{1/2}$ &  Final state &Decay type  & $Q_{0}$ (keV)& $Q^{*}_{0}$ (keV) & E$^{*}$  (keV)\\
\hline\noalign{\smallskip}
$^{76}$As (2$^{-}$)&26.24(9) h & $^{76}$Se (\{2$^{-}$, 3$^{-}$, 4$^{-}$\})&   $\beta^{-}$:  Allowed\{?\} &2960.6(9)&-7.8(11)   &2968.4(7)&\\
$^{155}$Tb (3/2$^{+}$)&  5.32(6) dy  & $^{155}$Gd (\{3/2\}$^{+}$) &EC: Allowed\{?\} & 820(10)&4(10)  &815.731(3)  &\\
   \end{tabular*}
   \label{table:low-Q1}
   \end{ruledtabular}
\end{table*}
\section{Experimental method}
\begin{figure}[!htb]
    \centerline{}
   \includegraphics[width=0.99\columnwidth]{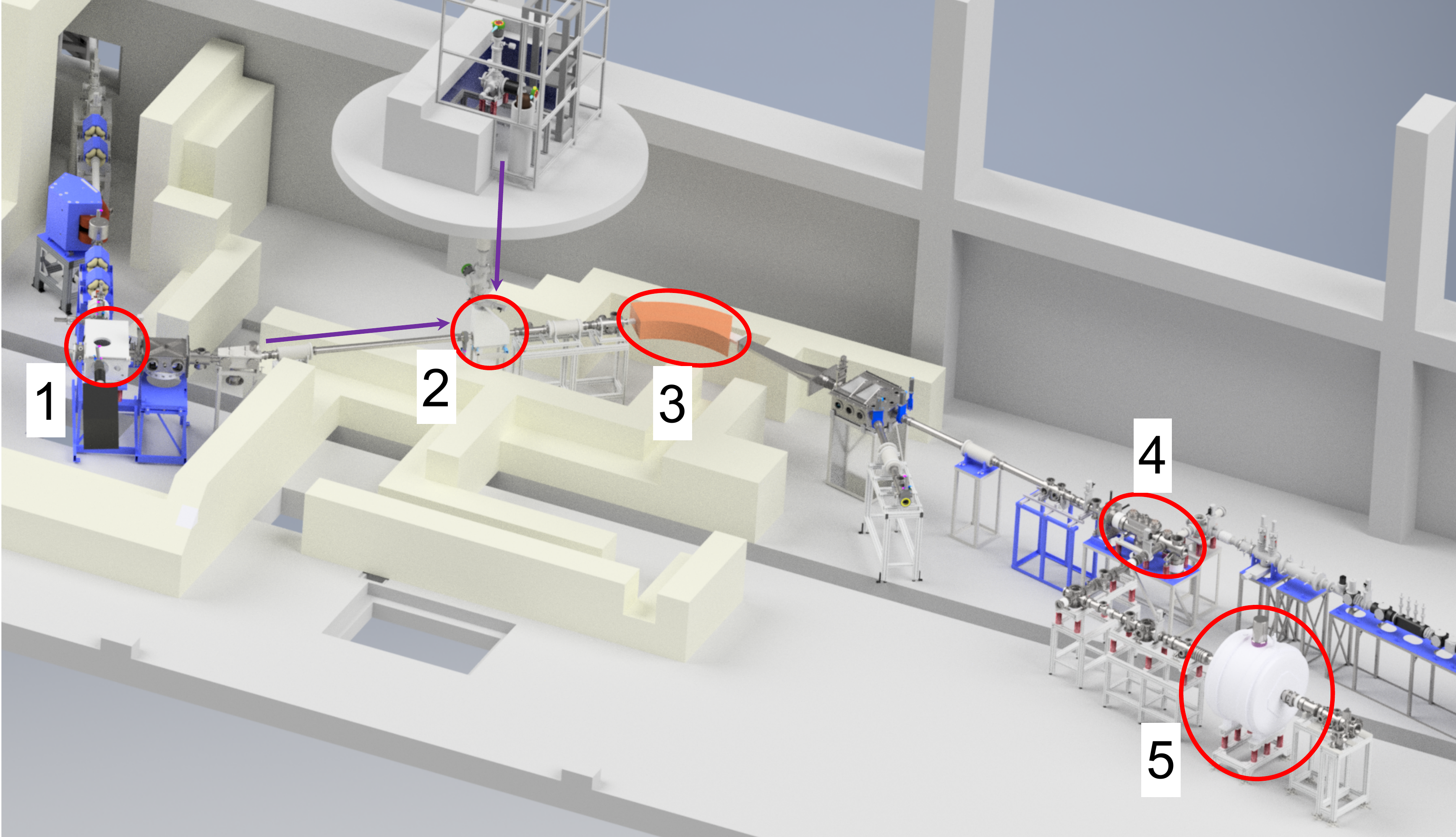}
   \caption{(Color online)   Schematic overview of the IGISOL facility. The ions were produced at the IGISOL target  chamber (1).  $^{76}$As$^{+}$ and $^{76}$Se$^{+}$ were produced with deuteron-induced  fusion reactions, while $^{155}$Tb$^{+}$ and $^{155}$Gd$^{+}$ were produced with proton-induced  fusion reactions.  After production and extraction from the gas cell, the ions were guided through an electrostatic bender (2),  mass number selected  with a dipole magnet (3), cooled and bunched in the RFQ cooler-buncher (4) and finally injected to the JYFLTRAP double Penning trap setup (5) for the mass-difference measurement . }
   \label{fig:igisol}
\end{figure}

The experiments  for the direct $Q_{0}$ value determination were carried out at the Ion Guide Isotope Separator On-Line facility (IGISOL) using the JYFLTRAP double Penning trap mass spectrometer~\cite{Eronen2012}, located at the University of Jyv\"askyl\"a~\cite{Moore2013,Kolhinen2013}. A schematic layout of the facility is shown in Fig.~\ref{fig:igisol}. For $\beta^-$ decay $^{76}$As $\rightarrow$ $^{76}$Se, the $Q_{\beta^-}$ value was deduced by measuring the cyclotron frequency ratio of singly charged $^{76}$As$^+$ and $^{76}$Se$^+$ ions. Similarly, for EC decay $^{155}$Tb $\rightarrow$ $^{155}$Gd, the $Q_{EC}$ value was derived from the cyclotron frequency ratio measurements of $^{155}$Tb$^+$ and $^{155}$Gd$^+$ ions.

To produce the $^{76}$As$^+$ and $^{76}$Se$^+$ ions simultaneously at IGISOL, a thin germanium target with a thickness of about 2~mg/cm$^2$ was bombarded with a 9-MeV deuteron beam from the K-130 cyclotron. The $^{155}$Tb$^+$ and $^{155}$Gd$^+$ ions were simultaneously produced with a 60-MeV proton beam bombarding a gadolinium target of about 2~mg/cm$^2$ in thickness.

After production, the ions were stopped and thermalized inside the gas cell of the IGISOL light-ion ion guide~\cite{Huikari2004}. Therein, they were transported by means of helium gas flow (typical gas pressure of $\sim$100 mbar)  through  an extraction nozzle to the sextupole ion guide (SPIG)~\cite{Karvonen2008} as a continuous ion beam. After extraction from the SPIG to a high-vacuum region, the ions were accelerated with a 30~kV potential. The ion beam was subsequently steered with an electrostatic kicker and  guided through a magnetic dipole mass separator. The separator was sufficient to reject all non-isobaric contaminants but  ions which were singly charged with a mass number of interest ($A$ = 76 or 155). The separated and selected ions were then transported through an electrostatic switchyard, which also houses a fast kicker electrode that was used to chop the beam to have an optimum number of ions. After the switchyard,  a radiofrequency quadrupole (RFQ) cooler-buncher~\cite{Nieminen2001} was used to cool and bunch the injected ions. Finally, the bunched and cooled ions were transported to the downstream JYFLTRAP double Penning trap for the frequency ratio measurement to determine the $Q$ values.

The JYFLTRAP double Penning trap consists of two cylindrical traps located in a 7-T superconducting  magnet.  Inside the traps, the homogeneous magnetic field and a quadrupolar electrostatic potential  are used to confine the cooled and bunched ions from the RFQ. The first trap is the purification trap, which is used for purification and separation of ions of interest. While the second trap, the  so-called precision trap, is utilized for a high-precision mass determination or a direct $Q$ value measurement. Two techniques are typically used, of which one is the conventional time-of-flight ion-cyclotron-resonance (TOF-ICR) method~\cite{Koenig1995,Graeff1980}, and another the phase-imaging ion-cyclotron-resonance (PI-ICR) technique~\cite{Nesterenko2018,Eliseev2014,Eliseev2013}.

The ions with $A$ = 76 from the deuteron-induced secondary beam, which consisted of  $^{76}$As$^+$,  $^{76}$Ga$^+$ and $^{76}$Se$^+$, were purified via the mass-selective buffer gas cooling method~\cite{Savard1991} in the  first  trap. A typical resolving power $M/\Delta M \approx 10^{5}$ was enough to remove all neighbouring isobaric ions and any other ion species present in the beam, resulting in a clean sample of $^{76}$As$^+$ or $^{76}$Se$^+$. 
%
Ions of $A$ = 155 from the proton-induced secondary beam were injected to JYFLTRAP, consisting of $^{155}$Tb$^+$, $^{155}$Gd$^+$, $^{155}$Sm$^+$ and $^{155}$Eu$^+$. These were firstly purified via the mass-selective buffer gas cooling method. An additional higher resolving power  Ramsey cleaning method~\cite{Eronen2008a} was utilized to selectively prepare a clean sample of $^{155}$Tb$^+$ or $^{155}$Gd$^+$. 

After the well-prepared clean ion sample of either the decay parent or decay daughter entered the second trap, the measurement of cyclotron frequency
\begin{equation}
\nu_{c}=\frac{1}{2\pi}\frac{q}{m}B
\label{eq:nuc}
\end{equation} 
 was performed. Here $q/m$ is the charge-to-mass ratio of the stored ion and $B$ the magnetic field strength.

In this work, the PI-ICR technique was used to measure the cyclotron frequencies~\cite{Nesterenko2018,Eliseev2014,Eliseev2013}, a method which is about 25 times faster reaching a certain precision compared to the TOF-ICR method. 
In particular, the measurement scheme number 2 described in~\cite{Eliseev2014} was utilized to directly measure the cyclotron frequency.
Two timing patterns, one called ``magnetron'' and the other ``cyclotron'' pattern were used, see ~\cite{Nesterenko2018}. These patterns are otherwise identical except for the switching-on instant of the $\pi$-pulse that converts ions' cyclotron motion to magnetron. In the ``magnetron'' pattern the ions revolve in the trap for an accumulation time $t$  with magnetron motion while in the ``cyclotron'' pattern the ions revolve with cyclotron motion. The exact knowledge of the switch-on time difference $t$ is essential. The used patterns produce so-called magnetron and cyclotron spots or phases on the position-sensitive micro-channel plate (MCP) detector~\cite{PS-MCP}.
%
Additionally, it is necessary to measure the motional center spot. With these data, it is then possible to get the angle between the magnetron and cyclotron motion phases with respect to the center spot
\begin{equation}\label{eq:alphac}
   \alpha_c = \alpha_+-\alpha_-,
\end{equation}
where $\alpha_+$ and $\alpha_-$ are the polar angles of cyclotron and magnetron phases, respectively. Finally, the cyclotron frequency $\nu_{c}$  is deduced from
\begin{equation}
\label{eq:nuc2}
\nu_{c}=\frac{\alpha_{c}+2\pi n_{c}}{2\pi{t}},
\end{equation}
where  $n_{c}$ is the number of complete revolutions during the phase accumulation time $t$.
The measurement is set up so that $\alpha_c$ will be small in order to minimize systematic shifts due to image distortion by choosing $t$ to be as close to integer-multiples of the $\nu_c$ period as possible.

The accumulation time $t$ for both $^{76}$Ga$^+$ and $^{76}$As$^+$ ions during the interleaved measurement was chosen to be 310~ms in order to ensure that the cyclotron spot was not overlapping with any possible isobaric, excited or molecular contamination. For both $^{155}$Tb$^{+}$ and $^{155}$Gd$^{+}$ ions, the accumulation time $t$ was tuned to be 620~ms.
The accumulation time $t$ was also tuned to the nearest integer-multiples of period of $\nu_c$ to minimize the angle $\alpha_c$. This ensured minimal influence from the interconversion of magnetron and cyclotron motions and also minimized the image distortion shift  \cite{Eliseev2014,Kretzschmar2012c}. In these measurements $\alpha_c$ did not exceed a few degrees.

\begin{figure}[!htb]
   \includegraphics[width=0.99\columnwidth]{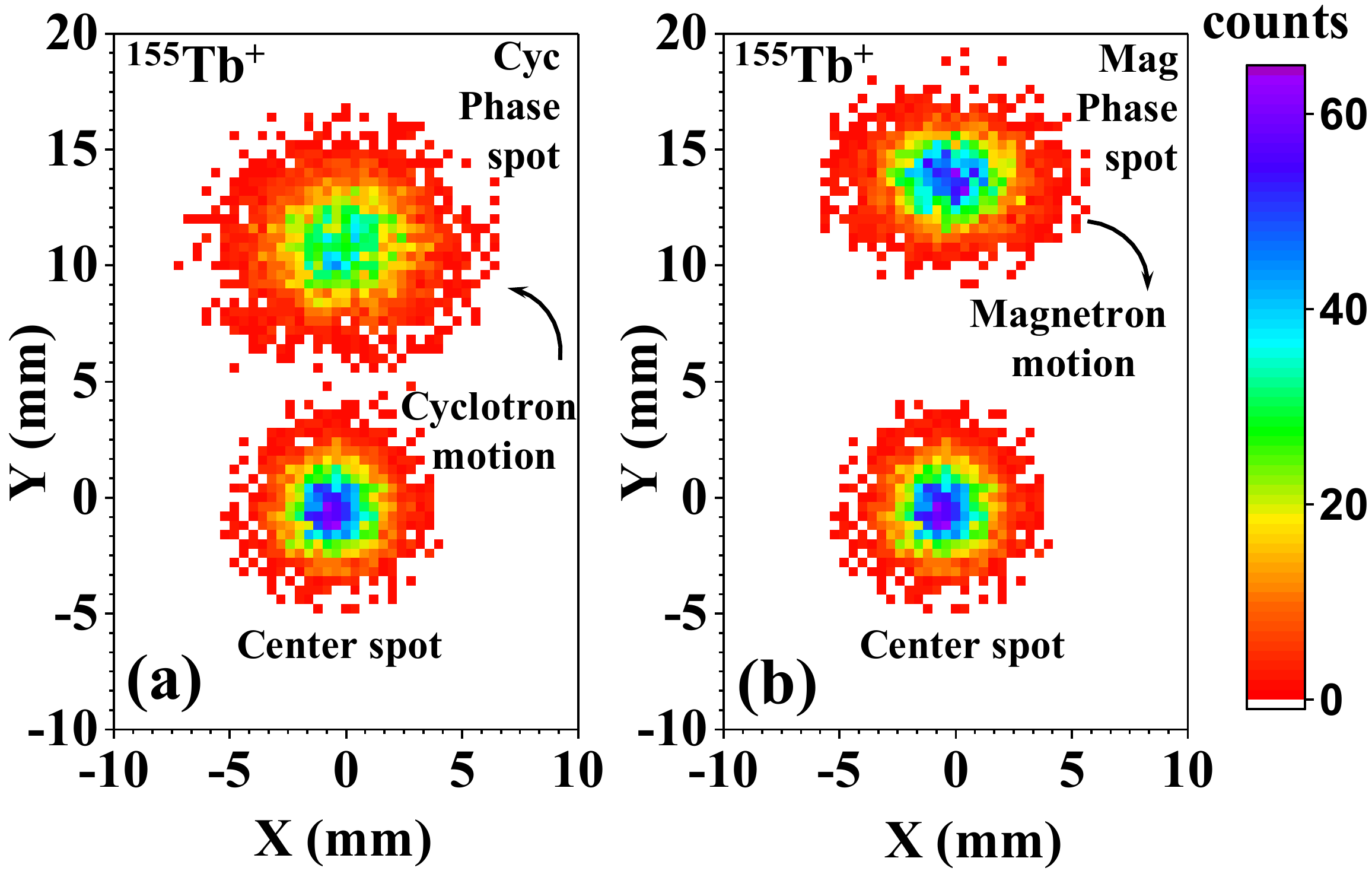}
   \caption{(Color online). Center, cyclotron phase and magnetron phase spots of $^{155}$Tb$^{+}$ on the position-sensitive MCP detector after the PI-ICR excitation pattern with an accumulation time of 620 ms composed of one set of data recorded in the experiment. The cyclotron phase spot is displayed on the left panel (a) and the magnetron phase spot with a center spot on the right (b). 
The color bar illustrates the number of ions in each pixel.}
   \label{fig:2-phases}
\end{figure}

The collected ``magnetron'' and ``cyclotron'' phase spots of $^{155}$Tb$^{+}$ ions are plotted in the left and right panels of Fig.~\ref{fig:2-phases}. The delay of the cyclotron motion excitation was repeatedly scanned over one magnetron period and the final extraction delay was varied over one cyclotron period to account for any residual magnetron and cyclotron motions that could shift the different spots. These constituted in total $5\times5=25$ scan points for both magnetron and cyclotron phase spots.

The decay $Q_{0}$ value is determined from the cyclotron frequency measurements in form of the ratio $R$:
\begin{equation}
\label{eq:Qec}
Q_{0}  = (R-1)(M_d - m_e)c^2+\Delta{B_{p,d}}.
\end{equation}
where $m_e$ is the electron mass and $\Delta B_{p,d}$ the electron binding energy difference of the parent-daughter atoms, which is typically on the order of few eV. 
The cyclotron frequency ratio is 
\begin{equation}
\label{eq:ratio}
    R = \frac{\nu_{c,d}}{\nu_{c,p}},
\end{equation}
where $\nu_{c,p}$ is the cyclotron frequency for parent $^{76}$As$^+$ and $\nu_{c,d}$ for daughter $^{76}$Se$^+$ measured in the deuteron-induced experiment, while $\nu_{c,p}$ is the cyclotron frequency for $^{155}$Tb$^+$ and $\nu_{c,d}$ for $^{155}$Gd$^+$ measured in the proton-induced experiment.
The ionization energies from the National Institute of Standards and Technology (NIST)~\cite{NIST_ASD} are used for the calculation of  $\Delta B_{m,d}$ in this work. For the case of parent $^{76}$As$^{+}$ and daughter $^{76}$Se$^{+}$, $\Delta B_{m,d}$ = -2.607(10) eV. While for the case of parent $^{155}$Tb$^{+}$ and daughter $^{155}$Gd$^{+}$,  $\Delta B_{m,d}$ = -0.563(28) eV.  By measuring mass doublets ($^{76}$As-$^{76}$Se or $^{155}$Tb-$^{155}$Gd), contributions to the measurement uncertainty arising from mass-dependent systematic effects due to frequency shifts, for example caused by field imperfections, become negligible compared to the typical statistical uncertainty achieved in the measurement.  Additionally, as the mass differences $\Delta M/M$ of parent-daughter for both $^{76}$As-$^{76}$Se and $^{155}$Tb-$^{155}$Gd are smaller than $10^{-4}$, the contribution to the $Q_{0}$ values from the mass uncertainties of the reference $^{76}$Se (0.08 keV/c$^2$) and $^{155}$Gd (1.1 keV/c$^2$) are negligible.
Due to the fact that the magnetic field changes with time, an accurate calibration of the field is required for a frequency ratio determination.
Thus, a linear interpolation of the neighbouring measurements of the cyclotron frequency of the daughter nuclide was performed to obtain the magnetic field  at the moment of the parent cyclotron frequency measurement. During these two experiments, we alternated between parent ion and daughter ion cyclotron frequency measurements every few minutes to minimize the contribution of the magnetic field fluctuation to the measured cyclotron frequency ratio. Therefore, nonlinear changes of the magnetic field between two neighbouring frequency measurements are negligible on the level of frequency determination uncertainties.

\section{Results and discussion}

The data were collected by a sequential measurement of the magnetron and cyclotron phase spots of ions of the parent nucleus and the daughter nucleus, followed by the center spot  recording with daughter-nucleus ions. 
Each of the steps lasted about six minutes.
For the case of $^{76}$As$^+$-$^{76}$Se$^+$, two separate runs were recorded, one lasting for 2.1 hours and the other for 5.3 hours and in total 7.4 hours. For the case of $^{155}$Tb$^+$-$^{155}$Gd$^+$, four runs were recorded. These lasted for 3.5 hours, 2.1 hours, 2.2 hours, and 1.7 hours for a total of 9.5 hours.

Before the determination of the positions of each spot, every four or eight rounds of data were grouped for analysis.
For each group, positions of each spot---magnetron, cyclotron and center---were determined using the maximum likelihood method and were used to calculate the phase angles. The calculated phase angles were finally used to deduce the cyclotron frequencies via Eq.~\ref{eq:nuc2}. 
The closest measured cyclotron frequencies of the daughter ions were linearly interpolated to the time of the measurement of parent ions in between. This interpolated frequency was used to deduce the cyclotron resonance frequency ratio $R$ via Eq.~\ref{eq:ratio}. The final ratio value is then calculated as the weighted  mean of all individual ratios.

During the measurement, the incident ion rate was intentionally limited to a maximum 5  ions/bunch with a median value of around 2 ions/bunch.  In order to reduce a possible cyclotron frequency shift due to ion-ion interactions~\cite{Kellerbauer2003,Roux2013}, only bunches with upto 5 ions or less  were considered in the analysis. A countrate-class analysis ~\cite{Kellerbauer2003},  an evaluation of the data with respect to the number of ions for given measurement cycles, was performed to confirm that  the frequency was indeed not shifting with respect to the number of ions. 

\begin{figure}[!htb]
   \includegraphics[width=0.99\columnwidth]{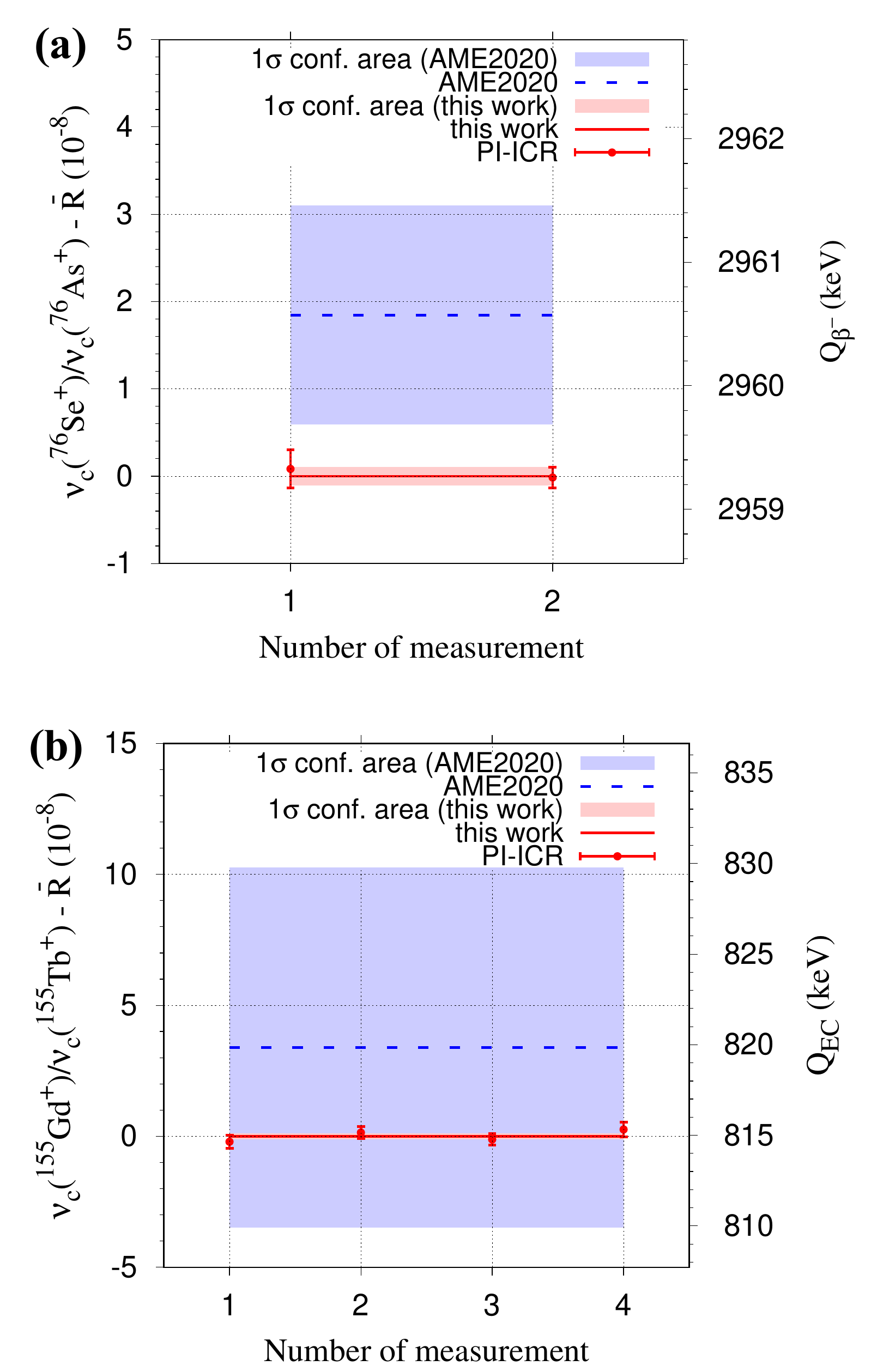}
   \caption{(Color online). Comparison of results from this work and AME2020 for  (a) $^{76}$As$^{+}$-$^{76}$Se$^{+}$ and (b) $^{155}$Tb$^{+}$-$^{155}$Gd$^{+}$ measurements. The left axis shows the frequency ratio deviation from the measured value and the right axis the corresponding $Q$ value. The red points are the data points and the solid horizontal red line with the shaded area ($1\sigma$ uncertainty band) the final value. The dashed blue line is the value derived from AME2020 (shaded area is the $1\sigma$ uncertainty band). 
   }
   \label{fig:ratio}
\end{figure}

As the $\alpha_c$ of Eq. (\ref{eq:alphac}) was kept as small as possible (typically below a few degrees) in the measurement, the frequency shifts due to ion image distortions were well below the statistical uncertainty. All individual ratios for the corresponding runs were calculated along with the inner and outer errors in the analysis of both $^{76}$As$^{+}$-$^{76}$Se$^{+}$ and $^{155}$Tb$^{+}$-$^{155}$Gd$^{+}$ measurements.  The Birge ratio~\cite{Birge1932}, the ratio of inner and outer errors, was calculated for each data set. The larger of the errors was taken as the final uncertainty. The final ratio $\overline{R}$  was calculated as the weighted mean of all ratios of the data sets for the PI-ICR data.  Table~\ref{table:Rvalue} lists the obtained frequency ratios $R$ of all runs and the final weighted mean ratio $\overline{R}$ of both measurements.  
The final frequency ratio $\overline{R}$ and the resulting $Q_{\beta^-}$ value are $1.000 041 846 2(10)$  and $2959.265(74)$~keV for $^{76}$As$^{+}$-$^{76}$Se$^{+}$, respectively. $\overline{R}$ and the $Q_{EC}$ value are $1.000 005 647 2(12)$  and $814.94(18)$~keV for $^{155}$Tb$^{+}$-$^{155}$Gd$^{+}$, respectively.  The final results of the analysis compared to literature values are also demonstrated in Fig.~\ref{fig:ratio}.

The reliability of the above interpolation method has been cross checked by a polynomial fitting method~\cite{Nesterenko2018} to deduce the frequency ratio. The frequency ratio obtained from the polynomial fit agrees well with the value from the linear interpolation analysis  within a combined 1$\sigma$ uncertainty for both $^{76}$As$^{+}$-$^{76}$Se$^{+}$ and $^{155}$Tb$^{+}$-$^{155}$Gd$^{+}$ measurements.

\begin{table}[!htb]
\caption{Cyclotron frequency ratios $R$ for each run and average values $\overline{R}$ for each isobaric ion pair. $N$ is the number of the run.}
\begin{ruledtabular}
   \begin{tabular*}{\textwidth}{ccccc}
Isobaric pair& $N$ & $R$ &$\overline{R}$  \\
\hline\noalign{\smallskip}
   \multicolumn{1}{c}{\multirow{2}{*}{$^{76}$As$^+$/$^{76}$Se$^+$}}&1&1.000 041 847 5(22)&{\multirow{2}{*}{1.000 041 846 2(10)}} \\
    \multicolumn{1}{c}{}&2&1.000 041 845 8(12)& \\
\hline\noalign{\smallskip}
   \multicolumn{1}{c}{\multirow{4}{*}{$^{155}$Tb$^+$/$^{155}$Gd$^+$}}&1&1.000 005 645 1(25)&{\multirow{4}{*}{1.000 005 647 2(12)}}\\
    \multicolumn{1}{c}{}&2&1.000 005 648 7(23)& \\
     \multicolumn{1}{c}{}&3&1.000 005 646 0(22)& \\
      \multicolumn{1}{c}{}&4&1.000 005 649 8(28)&  \\    
   \end{tabular*}
   \label{table:Rvalue}
\end{ruledtabular}
\end{table}

\begin{table}[!htb]
\caption{$Q_{0}$ value and the mass-excess (ME) determined in this work in comparison to the AME2020 values~\cite{Huang2021}.}
\begin{ruledtabular}
   \begin{tabular*}{\textwidth}{lcc}
& $Q_{0}$ (keV)& ME (keV/$c^2$) \\
\hline\noalign{\smallskip}
$^{76}$As (This Work)&   2959.265(74)& -72292.683(75) \\
$^{76}$As (AME2020) &     2960.6(9)&  -72291.4(9) \\
\hline\noalign{\smallskip}
$^{155}$Tb (This Work)& 814.94(18)  &   -71254.9(12) \\
$^{155}$Tb (AME2020) &  820(10)  &  -71250(10)
   \end{tabular*}
   \label{table:Qvalue}
\end{ruledtabular}
\end{table}

\begin{figure*}[!htb]
   \includegraphics[width=1.5\columnwidth]{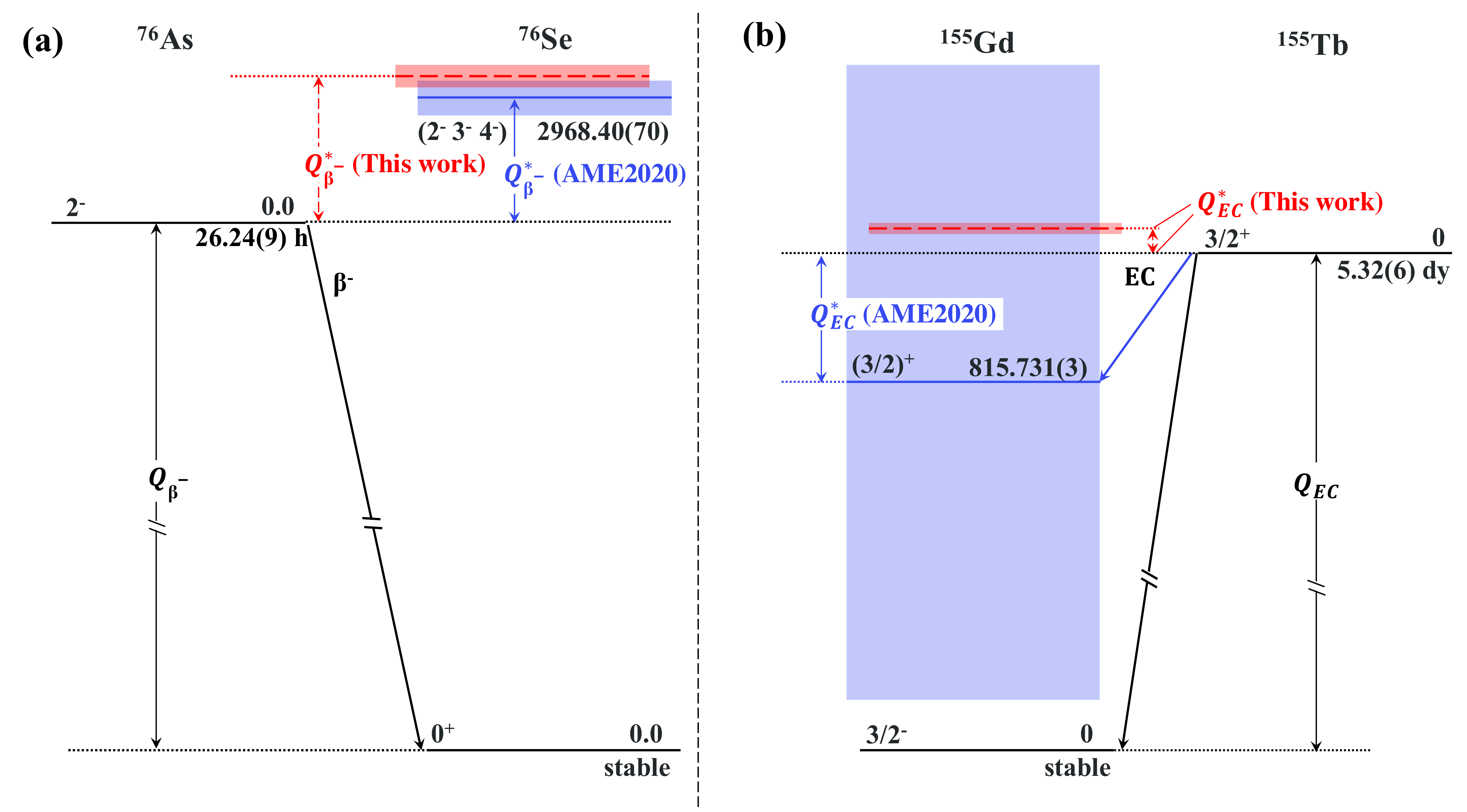}
   	\caption{(color online) Partial decay scheme of (a) $\beta^-$ decay of the $^{76}$As ground state to excited states in $^{76}$Se and (b) EC decay of the $^{155}$Tb ground state to excited states in $^{155}$Gd. 
   	$Q_{\beta^-}$/$Q_{EC}$ for both decays is the  ground-state-to-ground-state $Q$ value.
   	The energies of the excited states are taken from~\cite{NNDC}. The uncertain spin-parities are denoted with parentheses. The shaded bands in blue show the $1\sigma$ uncertainty in the $Q$ value as derived from the AME2020. The shaded bands in red show the $1\sigma$ uncertainty in the $Q$ value deduced from the newly determined $Q_{0}$ from this work. The decay $Q$ values and excitation energies are in units of keV. See also Table~\ref{table:low-Q1}, Table~\ref{table:Qvalue} and Table~\ref{table:low-Q}. 
}
   \label{fig:Q value-comparison}
\end{figure*}

The final $Q_{0}$ values and the mass-excess values of $^{76}$As and $^{155}$Tb obtained from the mean cyclotron frequency ratio measurements are listed in  Table~\ref{table:Qvalue}. 
In AME2020, the $Q_{0}$ value of $^{76}$As($\beta^{-}$)$^{76}$Se is deduced from the mass difference of $^{76}$As and $^{76}$Se. $^{76}$Se was measured directly with high precision via PTMS~\cite{Wang2021}. $^{76}$As mass is linked to the $^{75}$As(n,$\gamma$)$^{76}$As reaction measurement with an influence of 100\%. $^{75}$As mass was tied to $^{75}$As(p,n)$^{75}$Se and $^{75}$Se(p,$\alpha$)$^{75}$Se measurements with influences of 85.3\% and 14.7\%, respectively. These two reaction measurements give an evaluated uncertainty of 0.9 keV/$c^2$ for the $^{75}$As mass, which is the main contribution to the uncertainty of the $Q_{0}$ value of $^{76}$As($\beta^{-}$)$^{76}$Se. 
In AME2020, the $Q_{0}$ value  of $^{155}$Tb(EC)$^{155}$Gd is evaluated from the mass difference of $^{155}$Tb and $^{155}$Gd. 
The atomic mass of $^{155}$Gd is deduced from three independent measurements $^{155}$Gd((n,$\gamma$)$^{156}$Gd (70.1\% influence), $^{154}$Gd(n,$\gamma$)$^{155}$Gd (19.7\%  influence) and  $^{155}$GdO-C$_{15}$ (7.3\%  influence).  An uncertainty of 1.2 keV for  mass of $^{155}$Gd is given in AME2020. 
The atomic mass of $^{155}$Tb is primarily linked to $^{155}$Dy($\beta^{+}$)$^{155}$Tb. The $^{155}$Dy mass is related to $^{156}$Dy(d,t)$^{155}$Dy (92.1\%  influence) and $^{155}$Ho($\beta^{+}$)$^{155}$Dy (7.9\%  influence). $^{156}$Dy mass is primarily determined by PTMS (99.3\%  influence) and $^{155}$Ho mass is linked to $^{155}$Ho($\beta^{+}$)$^{155}$Dy (60.9\%  influence) and partially (39.1\%  influence) evaluated from the directly determined mass with a large uncertainty of 30 keV~\cite{Litvinov2005} via the storage ring using the time-resolved Schottky technique at GSI. 
Previous measurements show that the $Q$ values determined with an indirect method may possibly be incorrect, and even deviate by more than 10 keV from high-precision PTMS measurements
~\cite{Fink2012,Nesterenko2019,Ge2021}. 

As the $Q_{0}$ values of both decay pairs are linked to indirect measurements, direct measurements of the $Q_{0}$ values are strongly encouraged to confirm the accuracy with high precision.
For the decay pair $^{76}$As-$^{76}$Se, the $Q_{0}$ value from this work is a factor of 12 more precise and 1.33(90)~keV smaller than the value in AME2020~\cite{Huang2021}. 
For the decay pair $^{155}$Tb-$^{155}$Gd, the  $Q_{0}$ value from this work is a factor of 57 more precise and 5(10)~keV smaller than the value in AME2020~\cite{Huang2021}. 
The mass-excess values of $^{76}$As and $^{155}$Tb are improved by a factor of 12 and 8, respectively. The main contribution to the uncertainty in $^{76}$As is from the mass uncertainty (1.1~keV/c$^2$) of the reference daughter $^{76}$Se. For $^{155}$Tb, an additional 0.08~keV/c$^2$ uncertainty contribution to the mass-excess is added from the mass uncertainty of the reference daughter $^{155}$Gd.

\begin{table}[!htb]
   \caption{$Q$ values for the studied decays to the excited states of the daughter nuclei $^{76}$Se  and $^{155}$Gd. 
   The first column lists the daughter nucleus. The second column gives the experimental excitation energy $E^{*}$~\cite{NNDC} of the daughter state. The third and fourth columns give the $Q^*_{0}$ using the $Q_{0}$ from AME2020~\cite{Huang2021,Wang2021} and from this work, respectively. The last column shows the confidence ($\sigma$) of the $Q^*_{0}$ being negative.}
  \begin{ruledtabular}
   \begin{tabular*}{\textwidth}{ccccc}
Daught &E$^{*}$& \makecell[c]{$Q^*_{0}$ (keV) \\ (AME2020)} &\makecell[c] {$Q^*_{0}$ (keV) \\(This work)}& \makecell[c]{$Q/\delta Q$ \\(This work)}  \\
\hline\noalign{\smallskip}
$^{76}$Se & 2968.4(7)     &-7.8(11)  &-9.13(70) & 13 \\ 
$^{155}$Gd  & 815.731(3)   &4(10)  &-0.79(18) & 4 \\
   \end{tabular*}
   \label{table:low-Q}
   \end{ruledtabular}
\end{table}

With the direct measurements of the $Q_{0}$ values combined with the excitation energies from ~\cite{NNDC}, the $Q_{0}^*$ values of the potential ultra-low $Q$-value decays from ground states to the excited states are derived and tabulated in table~\ref{table:low-Q}. A comparison of  $Q$ values derived from AME2020 with the newly determined $Q$ values of this work is depicted in Fig.~\ref{fig:Q value-comparison} as well.  Our results confirm that the $\beta^{-}$ decay of the ground-state of $^{76}$As to the excited-state of $^{76}$Se (2968.4(7) keV)  is an energetically-forbidden decay branch with a $Q_{0}^*$ value below 0 at a 13$\sigma$ level. In addition, the EC decay of the ground-state of $^{155}$Tb to the excited-state of $^{155}$Gd (815.731(3) keV) is verified to be energetically forbidden at a level of 4$\sigma$.

While the small negative $Q$ values exclude these nuclei as potential candidates for neutrino-mass measurements, they make studies of an interesting rare process, a radiative detour transition, possible. This second-order process, involving an additional photon, can have a non-negligible branching ratio when angular-momentum selection rules  hinder the direct transition \cite{Longmire49}. In Ref. \cite{Pfutzner2015} the contribution of the virtual transition was estimated to be about 4\% of the $\beta$-$\gamma$ decay in $^{59}$Ni. However, the negative $Q$-value there was 26 keV, and the probability of the virtual transition is proportional to the inverse of the square of this energy difference \cite{Longmire49}. Thus, the smaller the energy difference, the better the candidate. Recently, another possible candidate, $^{72}$As, was discovered with a $Q$-value of -3.42(8) keV for a transition to an excited state \cite{Ge2021}. 

\section{Conclusion and Outlook}
At the JYFLTRAP Penning-trap mass spectrometer, direct high-precision ground-state-to-ground-state  $\beta^{-}$/EC-decay $Q$-value measurements of $^{76}$As($\beta^{-}$)$^{76}$Se and  $^{155}$Tb(EC)$^{155}$Gd were performed using the PI-ICR technique. 
The $Q$ values of the two decay pairs were determined to be 2959.265(74) keV and 814.94(18) keV, respectively. The precision of the $^{76}$As($\beta^-$)$^{76}$Se-decay $Q$ value was improved by a factor of more than 50. A factor of 12 more precise $Q$ value was obtained for the $^{155}$Tb(EC)$^{155}$Gd decay. 
Combining with the high-precision energy-level data from $\gamma$-ray spectroscopy, we deduced with sub-keV precision the  $Q$ values for the potential ultra-low $Q$-value $\beta^-$ transition $^{76}$As (ground state) $\rightarrow$ $^{76}$Se$^*$ (2968.4(7) keV) and EC transition $^{155}$Tb (ground state) $\rightarrow$ $^{155}$Gd$^*$ (815.731(3) keV) to be -9.13(70) keV and -0.79(18) keV, respectively. Both of the $\beta^{-}$ decay and EC decay candidate transitions were confirmed to be energetically forbidden at least at the level of 4$\sigma$, thus definitely ruling out these two decay transitions as possible ultra-low $Q$-value candidates for the electron (anti)neutrino mass determination. On the other hand, the discovery of such small negative $Q$ values makes these nuclei excellent candidates for the study of virtual $\beta$-$\gamma$ transitions.

\acknowledgments 
We acknowledge the staff of the accelerator laboratory of University of Jyv\"askyl\"a (JYFL-ACCLAB) for providing stable online beam and J.~Jaatinen and R.~Sepp\"al\"a for preparing the production targets. We thank the support by the Academy of Finland under the Finnish Centre of Excellence Programme 2012-2017 (Nuclear and Accelerator Based Physics Research at JYFL) and projects No. 306980, 312544, 275389, 284516, 295207, 314733, 318043, 327629 and 320062. The support by the EU Horizon 2020 research and innovation program under grant No. 771036 (ERC CoG MAIDEN) is acknowledged.
 

\bibliographystyle{apsrev4-1}

\bibliography{my-final-bib-from-jabref}
\end{document}